\documentclass[aps,twocolumn,nofootinbib,notitlepage,showkeys,superscriptaddress,preprintnumbers]{revtex4-2}

\usepackage{graphicx}
\usepackage{dcolumn}
\usepackage{bm}

\usepackage{amsmath,amssymb}
\usepackage{float}
\usepackage{multirow}
\usepackage{slashed}
\usepackage{xcolor}
\usepackage{physics}
\usepackage{multirow}
\usepackage[colorlinks=true, pdfstartview=FitV, bookmarks=true, bookmarksnumbered=true, breaklinks]{hyperref}

\usepackage{mathtools,braket}
\usepackage{soul}
\usepackage{lipsum}  
\usepackage{color}
\definecolor{blue}{rgb}{0.0, 0.0, 1.0}
\definecolor{red}{rgb}{1.0, 0.0, 0.0}
\definecolor{royalblue}{rgb}{0.0, 0.14, 0.4}
\hypersetup{linkcolor=royalblue, citecolor=blue, urlcolor=royalblue}

\usepackage{hyperref}
\hypersetup{colorlinks=true,citecolor=blue,linkcolor=blue,urlcolor=blue}

\usepackage[mathlines]{lineno}

\usepackage{newtxmath}
\usepackage{mathrsfs}

\usepackage[normalem]{ulem}
\usepackage{color}

\usepackage{tikz,xcolor,hyperref}
\definecolor{lime}{HTML}{A6CE39}
\DeclareRobustCommand{\orcidicon}{%
	\begin{tikzpicture}
	\draw[lime, fill=lime] (0,0) 
	circle [radius=0.16] 
	node[white] {{\fontfamily{qag}\selectfont \tiny ID}};
	\draw[white, fill=white] (-0.0625,0.095) 
	circle [radius=0.007];
	\end{tikzpicture}
	\hspace{-2mm}
}

\foreach \x in {A, ..., Z}{%
	\expandafter\xdef\csname orcid\x\endcsname{\noexpand\href{https://orcid.org/\csname orcidauthor\x\endcsname}{\noexpand\orcidicon}}
}

\begin{document}
\title{Inclusive Neutral-Kaon Photoproduction on the Deuteron}

\author{Mutoharoh\orcidA{}}
\email[E-mail: ]{mutoharoh41@ui.ac.id} %
\affiliation{Departemen Fisika, FMIPA, Universitas Indonesia, Depok 16424, Indonesia}

\author{Agus Salam\orcidB{}} 
\email[E-mail: ]{agus.salam@sci.ui.ac.id}
\affiliation{Departemen Fisika, FMIPA, Universitas Indonesia, Depok 16424, Indonesia}

\author{Terry Mart\orcidC{}}
\email[E-mail: ]{terry.mart@sci.ui.ac.id}
\affiliation{Departemen Fisika, FMIPA, Universitas Indonesia, Depok 16424, Indonesia}
\date{\today}

\begin{abstract}
We report on our study of inclusive neutral-kaon photoproduction on the deuteron, $d(\gamma,K^0)YN$, for photon energies between 0.9 and 1.1 GeV. The calculation is performed in the impulse approximation, with the deuteron wave function generated from the Bonn One-Boson-Exchange-Potential in $q$-space (OBEPQ) within a non-relativistic framework, while the kinematics and the elementary operator are kept relativistic. For the elementary operator we employ a recently developed isobar model that includes high-spin nucleon and $\Delta$ resonances and has been constrained by nearly 20,000 data points, and we compare its predictions systematically with those of the Kaon-Maid model. For the first time, the NKS1 and NKS2 data are subjected to a quantitative analysis within a modern elementary production framework. The New Operator provides a considerably better description of the data, whereas Kaon-Maid overestimates the measured cross section by up to a factor of four at the higher photon energies, a discrepancy that can be traced back to its unconstrained $\gamma n \to K^0\Lambda$ amplitude. We further show that the finite photon-energy and kaon-angle bins of the existing measurements generate a theoretical uncertainty comparable to the difference between the two models, so that a proper comparison with the data requires the calculation to be averaged over the experimental acceptance. A three-dimensional mapping of the cross section over the kaon momentum and angle reveals a narrow quasi-free ridge, accompanied by a second structure associated with the opening of the $\Sigma$ channels, from which we identify the kinematics most favorable for future measurements. Finally, the tensor target asymmetries are found to be far less sensitive to the elementary operator than the cross section, and therefore probe the nuclear dynamics in a way that is complementary to the cross section.
\end{abstract}


\maketitle

\section{Introduction} \label{sec:intro}

One of the central goals of hadronic physics is to understand the strong interaction in its non-perturbative regime, where the relevant degrees of freedom are hadrons rather than quarks and gluons. In this context, the extension of the well-constrained nucleon--nucleon ($NN$) interaction toward a unified description of the complete baryon--baryon interaction, as suggested by $SU(3)$ flavor symmetry, requires detailed knowledge of the hyperon--nucleon ($YN$) interaction. Direct $YN$ scattering experiments are notoriously difficult, since hyperon beams are short-lived and the available data base is small compared with that of the $NN$ system. Electromagnetic production of strangeness, in which a kaon and a hyperon are created simultaneously in order to conserve strangeness, therefore provides one of the few practical routes to this information. A quantitative description of such a reaction requires knowledge of both the electromagnetic ($\gamma NN$, $\gamma YY$, and $\gamma KK$) and the hadronic ($KYN$) interactions, which can be extracted from experimental data within phenomenological frameworks such as isobar models.

Research on kaon photoproduction on the nucleon has consequently remained a pivotal topic in hadronic physics, owing to its relevance for nucleon resonance dynamics and meson--baryon interactions. Early studies, such as the work by Thom in 1966, analyzed the reaction $\gamma p \rightarrow K^+ \Lambda$ through Feynman diagrams for the Born terms and partial-wave expansions for the resonances \cite{Thom:1966rm}. Subsequent developments incorporated fully diagrammatic approaches to systematically evaluate resonance contributions, laying the foundation for modern field-theoretic isobar models and, ultimately, for the construction of elementary production operators suitable for applications to kaon photoproduction on nuclear targets \cite{Adelseck:1985scp,Adelseck:1990ch}. Since then, considerable efforts have been devoted to this goal. Numerous calculations based on isobar models and multipole parameterizations \cite{Corthals:2005ce,deCruz:2011xi,Vrancx:2014pwa,Clymton:2017nvp,Mart:2017mwj,Skoupil:2018vdh} have been developed and subsequently refined as increasingly precise data have become available through advances in accelerator and detector technologies \cite{Mart:2026tyg}. However, only a small number of these models have included the $K^0$ production channels in their analyses \cite{David:1995pi,Maxwell:2015psa}. The datasets employed by the existing isobar models are summarized in Table 5 of Ref.~\cite{Mart:2026tyg}.

Because the kaon photoproduction thresholds lie substantially higher than those of pion photoproduction, a large number of baryon and meson resonances can already contribute near threshold. As a consequence, the predictive power of an isobar model is dictated almost entirely by the quantity and the quality of the data base used to constrain its couplings. The widely used Kaon-Maid model \cite{Mart:1999ed,Bennhold:1999mt,kaonmaid} was fitted to roughly 300 data points available at the end of the 1990s, whereas the New Operator of Ref.~\cite{Luthfiyah:2021yqe}, which includes nucleon and $\Delta$ resonances with spins up to $13/2$ formulated with consistent interactions, was constrained by nearly 20,000 data points measured at JLab, ELSA, MAMI, SPring-8, and ESRF. Comprehensive overviews of these developments can be found in Refs.~\cite{Mart:2025ufa,Mart:2026tyg}.

A major challenge in photoproduction studies is the lack of free-neutron targets, as free neutrons are unstable and have a relatively short lifetime. To overcome this difficulty, light nuclei such as deuterium ($d$ or $^2$H) and helium-3 ($^3$He) are commonly used as effective neutron targets, since each contains a bound neutron that can participate in the reaction. Among these systems, the deuteron is particularly advantageous due to its simple structure (one proton and one neutron) and its small binding energy ($\approx 2.2$ MeV), which allows a cleaner separation between the active nucleon and the spectator nucleon. For these reasons, the deuteron is widely regarded as an excellent target for investigating photoproduction processes in neutron channels \cite{Salam:2006kk}. The neutral-kaon channels are especially interesting in this respect. Since the $K^0$ carries no electric charge, the $t$-channel Born term is absent in the $\gamma n \to K^0\Lambda$ and $\gamma n \to K^0\Sigma^0$ processes, and isospin conservation forbids the $\Delta$ contribution in $\gamma n \to K^0\Lambda$. The neutral channels are therefore governed by a different combination of Born and resonance contributions than the extensively measured $\gamma p \to K^+\Lambda$ channel, and therefore, they provide a much more stringent test of the elementary operators than the proton channels alone.

The experimental relevance of these channels is illustrated by the studies of neutral-kaon photoproduction on the deuteron reported by Tsukada \textit{et al.} \cite{Tsukada:2007jy} (NKS1) and Futatsukawa \textit{et al.} \cite{Futatsukawa:2012zza} (NKS2) at the Laboratory of Nuclear Science in Sendai, which measured the inclusive $d(\gamma,K^0)YN$ cross section close to threshold, as well as by the later CLAS measurement of the $\gamma d \to K^0 \Lambda(p)$ reaction in the resonance region \cite{CLAS:2017gsu}. Despite the availability of these data for many years, they have not previously been subjected to a systematic quantitative analysis aimed at constraining the elementary production operator. Moreover, the measurements are unavoidably averaged over finite photon-energy and kaon-angle bins, a feature that must be properly accounted for before drawing conclusions about the elementary operator.

On the theoretical side, kaon photoproduction on the deuteron has been studied in considerable detail. Yamamura \textit{et al.} \cite{Yamamura:1999xm} evaluated the inclusive $K^+$ and exclusive $K^+Y$ processes with modern $YN$ forces and found sizable final-state interaction (FSI) effects near the $K^+\Lambda N$ and $K^+\Sigma N$ thresholds, in particular a cusp-like structure whose shape reflects the $S$-matrix pole position of the $\Lambda N$--$\Sigma N$ system. Salam and Arenh\"ovel \cite{Salam:2004gz} extended the analysis by including kaon--nucleon rescattering and the pion-mediated process $\gamma d \to \pi NN \to KYN$, and showed that these two-step mechanisms become important mainly at backward kaon angles, while their influence in the forward direction is small. Polarization observables were investigated by Miyagawa \textit{et al.} \cite{Miyagawa:2006kj}, who demonstrated that the double-polarization observable $C_z$ is particularly sensitive to the $YN$ interaction. Finally, Salam \textit{et al.} \cite{Salam:2006kk} analyzed the $K^0$ channels and showed that in quasi-free scattering (QFS) kinematics the FSI effects become negligible, so that the elementary amplitude can be extracted algebraically from the deuteron cross section. Importantly, they also found that, in this kinematic region, the uncertainty associated with the choice of elementary operator is larger than that arising from FSI. The deuteron thus acts not only as an effective neutron target, but also as a sensitive testing ground for the elementary production operator itself. These findings therefore suggest that, at forward kaon angles and under quasi-free kinematics, elementary observables such as the differential cross section may be extracted from kaon photoproduction on the deuteron with relatively small corrections from FSI. A related treatment is provided by the spectator approximation, in which the deuteron cross section is expressed in terms of the elementary differential cross section folded with the deuteron momentum distribution, together with the appropriate kinematic factors \cite{Bydzovsky:2004ev}. This formulation provides an alternative means of extracting the elementary production cross section from deuteron data. However, the accuracy of the spectator approximation and its applicability to the available experimental data need to be assessed quantitatively before drawing firm conclusions about its reliability for extracting elementary observables.

More recent investigations of the elementary processes adopt sophisticated isobar and effective Lagrangian frameworks, enabling the inclusion of high-spin nucleon and delta resonances while accounting for off-shell effects of bound nucleons in nuclear targets \cite{Luthfiyah:2021yqe}. In this context, we have recently examined kaon photoproduction on the deuteron within the impulse approximation, focusing on quasi-free kinematics and the influence of nuclear binding on the observable cross sections and polarization parameters. That study demonstrated that the off-shell contributions of the bound neutron are generally small in the inclusive cross sections and highlighted the differences between the forward and backward scattering configurations \cite{Mutoharoh:2026rxn}.

Based on these considerations, the present study investigates neutral kaon photoproduction on the deuteron under quasi-free kinematics with four specific objectives: (i) to confront the predictions of the New Operator and of Kaon-Maid with the available NKS1 and NKS2 inclusive data, (ii) to quantify the theoretical uncertainty induced by the finite photon-energy and kaon-angle bins of these measurements, (iii) to map the inclusive cross section over the full $p_K$--$\cos\theta$ plane in order to identify the kinematics most favorable for future experiments, and (iv) to explore the tensor target asymmetries as complementary observables. To this end, we focus on the kinematic region covered by the existing data and employ the formalism developed previously in Refs.~\cite{Salam:2004gz,Salam:2006kk}, taking Kaon-Maid~\cite{Mart:1999ed,Bennhold:1999mt,kaonmaid} and the New Operator~\cite{Luthfiyah:2021yqe} as the elementary production models that serve as inputs to the nuclear calculation.

The present paper is organized as follows. Section~\ref{sec:Theoretical_Framework} presents the theoretical framework, including the elementary operator employed in this study and the formalism for kaon photoproduction on the deuteron. The off-shell prescriptions adopted in our formalism are also discussed in this section. Section~\ref{sec:Results_and_Discussion} presents the numerical results and their discussion. Finally, Sect.~\ref{sec:Conclusion} presents the conclusions of this study.

\section {Theoretical Framework}
\label{sec:Theoretical_Framework}
\subsection{Elementary Operator}
\begin{figure*}[htbp]
\centering    \includegraphics[width=0.7\textwidth]{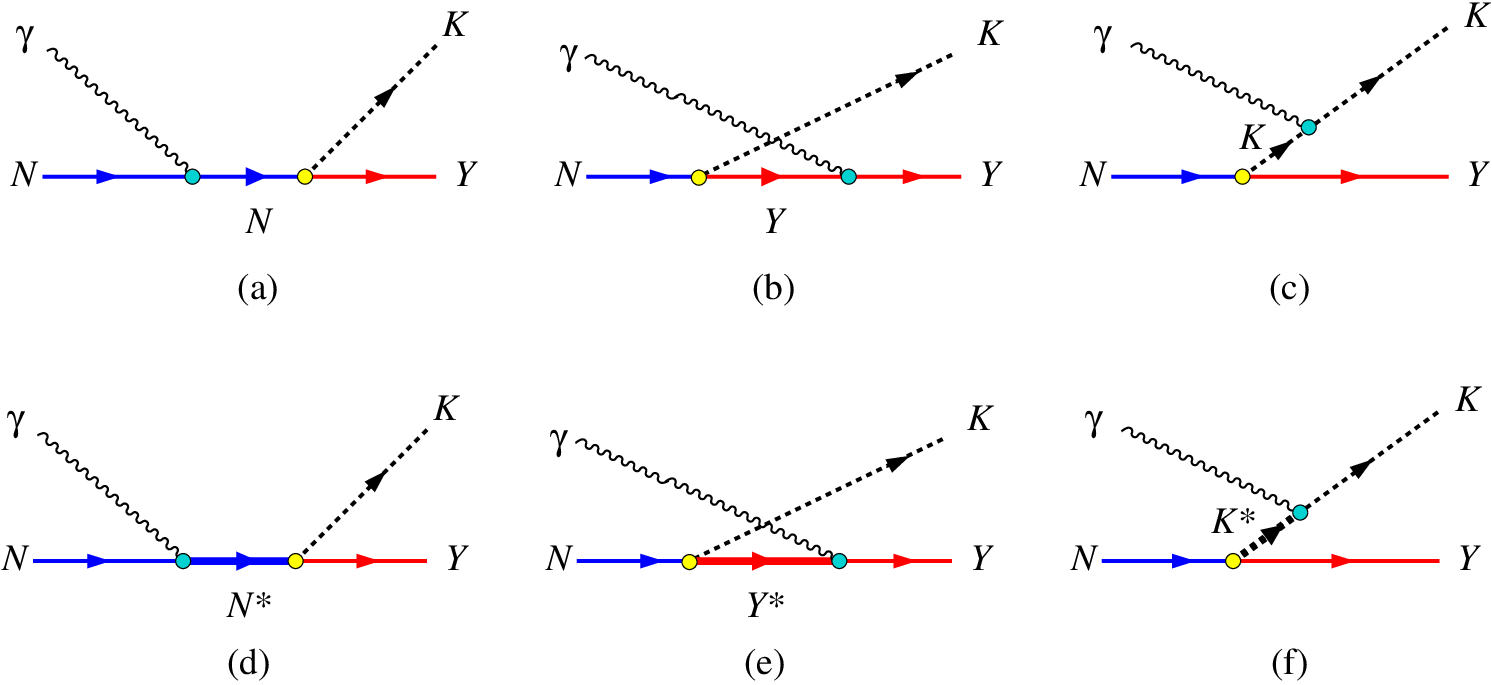} 
\caption{Feynman diagrams for kaon photoproduction on the nucleon. Panels~(a)--(c) illustrate the Born contributions, corresponding to the $s$-channel nucleon exchange, $u$-channel hyperon exchange, and $t$-channel kaon exchange, respectively. Panels~(d)--(f) show the resonance contributions involving intermediate excited baryon and meson states.}
\label{fig:diagram_Feynman_elementer}
\end{figure*}

The isobar model employed in the present work was developed in our previous studies \cite{Mart:1999ed,Luthfiyah:2021yqe} and presented in a compact form in our recent review~\cite{Mart:2025ufa,Mart:2026tyg}. Within this framework, the transition amplitude for kaon photoproduction on a nucleon is derived from an effective Lagrangian describing the electromagnetic coupling of the photon and the hadronic interactions between mesons and baryons. These interactions are represented by tree-level Feynman diagrams, as illustrated in Fig.~\ref{fig:diagram_Feynman_elementer}. The diagrams comprise three distinct reaction channels, namely the $s$-, $t$-, and $u$-channels, involving both ground-state hadrons and excited baryon resonances. Note that the diagrams shown in Fig.~\ref{fig:diagram_Feynman_elementer} are used for all six isospin channels of kaon photoproduction 

The amplitude required for calculating the cross sections in both the nucleon and nuclear cases was presented in Ref.~\cite{Mart:2025ufa}, where four alternative forms were discussed. For the convenience of the reader, we briefly review here the derivation of the amplitude most suitable for the present purpose. We adopt the following convention for the four-momenta of the incoming and outgoing particles in the elementary photoproduction process,
\begin{equation}
    \gamma(k)+N(p_N)\to K(q)+Y(p_Y) ,
\end{equation}
from which the Mandelstam variables are defined as
\begin{equation}
    s=(k+p_N)^2, \quad t=(k-q)^2, \quad {\rm and} \quad u=(k-p_Y)^2 ,
\end{equation}
with $k^2=0$. Using the standard Feynman diagrammatic technique, the total amplitude corresponding to the diagrams shown in Fig.~\ref{fig:diagram_Feynman_elementer} can be decomposed into four gauge- and Lorentz-invariant matrices $M_i$ and Lorentz-invariant scalar functions $A_i$ as 
\begin{equation}
    \label{eq:elementary_ampl_M}
    {t}^{\gamma K} = {\bar u}_Y \sum_{i=1}^{4} A_i(s,t,u)\, M_i\, u_N ,
\end{equation}
with \cite{Mart:2019mtq}
\begin{subequations}
\label{eq:Ai_Mi}
\begin{align}
M_1 ~&=~ \gamma_5\slashed{\epsilon}\slashed{k},\\
M_2 ~&=~ 2\gamma_5\left(q\cdot \epsilon\, P\cdot k-q\cdot k\, P\cdot \epsilon\right), \\
M_3 ~&=~ \gamma_5\left(q\cdot k\, \slashed{\epsilon}-q\cdot\epsilon\,\slashed{k}\right),\\
M_4 ~&=~ i\varepsilon_{\mu\nu\rho\sigma}\gamma^\mu q^\nu\epsilon^\rho k^\sigma,
\end{align}
\end{subequations}
where $\epsilon_\mu$ denotes the photon polarization vector, $P \equiv (p_N+p_Y)/2$, and $\varepsilon_{\mu\nu\rho\sigma}$ is the four-dimensional Levi-Civita tensor. 

In the elementary process, the hadronic coupling constants and other unknown parameters, such as the electromagnetic couplings of the resonances, are determined by fitting the calculated observables to experimental data. The Kaon-Maid model was constrained by a fit to approximately 300 data points, whereas the New Operator was developed using nearly 20,000 data points, including more recent measurements obtained with modern accelerator and detector facilities at Jefferson Lab (JLab) in Virginia, SPring-8 in Japan, ELSA in Bonn, MAMI in Mainz, and ESRF in Grenoble. The substantially larger and more recent data set used in constructing the New Operator enables it to provide a more comprehensive description of the available experimental data than the Kaon-Maid model. 

\begin{figure*}[t]
\centering \includegraphics[width=0.85\textwidth]{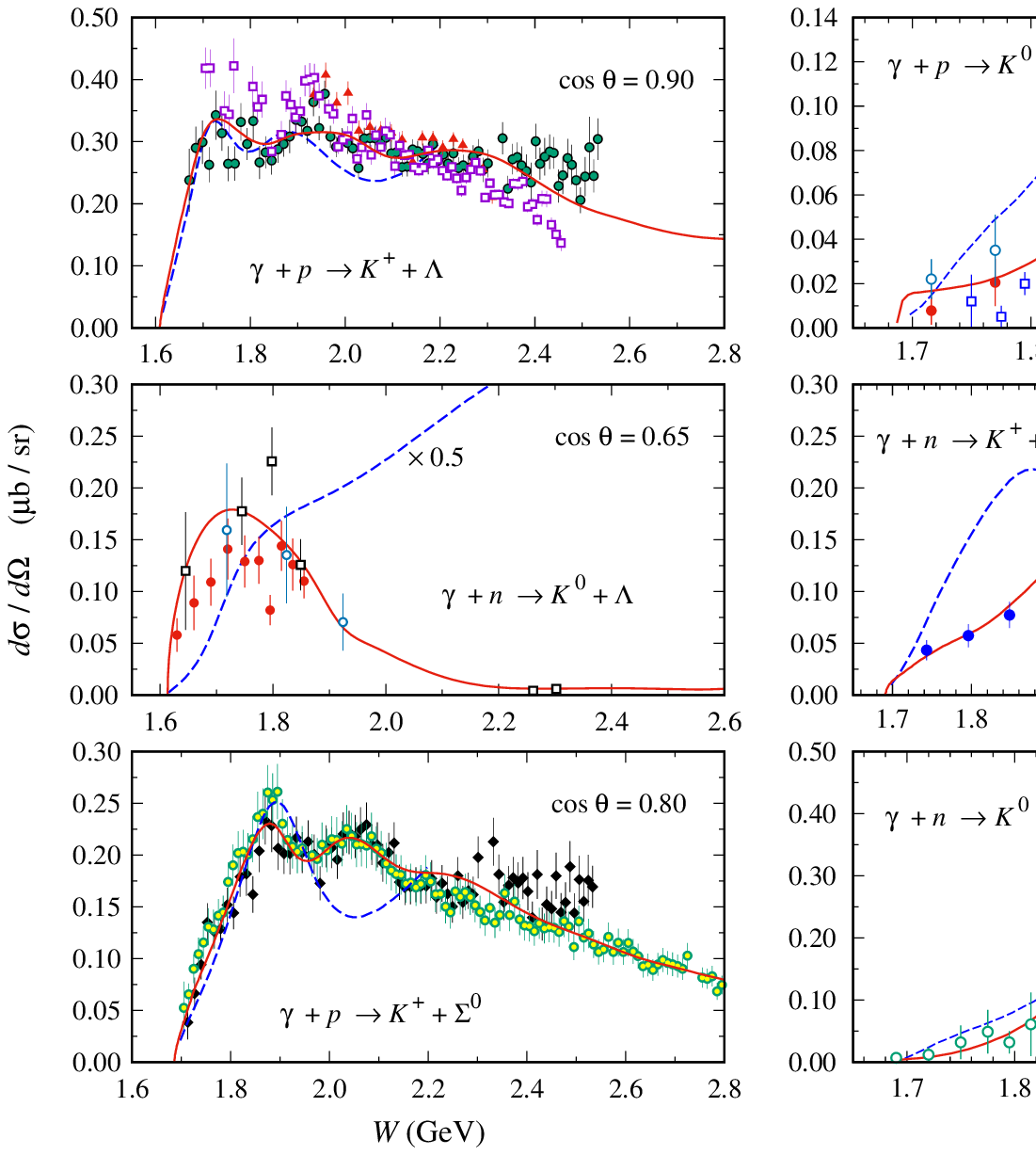} 
\caption{\label{fig:elementary_dcs} Samples of the differential cross sections for kaon photoproduction on the nucleon in six isospin channels, comparing the predictions of two theoretical models with experimental data. The dashed lines represent the predictions of Kaon-Maid~\cite{Mart:1999ed,Bennhold:1999mt,kaonmaid}, while the solid lines show the results obtained using the new operator introduced in Ref.~\cite{Luthfiyah:2021yqe}. The experimental data are taken from Refs.~\cite{CrystalBallatMAMI:2013iig,CLAS:2009rdi,CLAS:2005lui,LEPS:2005hji,CLAS:2017gsu,A2:2018doh,CLAS:2010aen,A2:2013cqk,Kohri:2006yx,CLAS:2009fmu}. See Ref.~\cite{Mart:2025ufa} for further details on the data included in this figure.}
\end{figure*}

Samples of the results obtained from the two phenomenological models are shown in Fig.~\ref{fig:elementary_dcs}, where the calculated differential cross sections are compared with the available experimental data ~\cite{CrystalBallatMAMI:2013iig,CLAS:2009rdi,CLAS:2005lui,LEPS:2005hji,CLAS:2017gsu,A2:2018doh,CLAS:2010aen,A2:2013cqk,Kohri:2006yx,CLAS:2009fmu} in the forward-angle region (small kaon angles $\theta$). This kinematic region is of particular interest because the nuclear cross sections are sizable only near the forward direction. Moreover, the available experimental data for kaon photoproduction on the deuteron are limited to $0.80 < \cos\theta < 1.00$. Within this kinematic range, the difference between the two phenomenological models is clearly visible: the New Operator exhibits substantially better agreement with the experimental data than the Kaon-Maid model. 

It is also important to note that a mutual inconsistency was identified between the SAPHIR~\cite{Glander:2003jw} and CLAS~\cite{Bradford:2005hq} data for the $\gamma+p\to K^++\Lambda$ channel \cite{Bydzovsky:2006wy}. For this reason, we exclude the SAPHIR data from our analysis \cite{Luthfiyah:2021yqe}. Even with the SAPHIR data excluded, the remaining experimental data are still relatively scattered, particularly in the vicinity of the threshold region. Nevertheless, the New Operator provides significantly better agreement with the experimental data than the Kaon-Maid model. Given the decisive role of the elementary production operator in the nuclear calculation, future measurements in this kinematic region would therefore be highly valuable.

Since the available data for neutral-kaon photoproduction on the deuteron are restricted to $E_\gamma < 1.2$ GeV, corresponding approximately to $W < 1.8$ GeV for the elementary process on a nucleon at rest, Fig.~\ref{fig:elementary_dcs} provides an indication of the expected difference between the nuclear cross sections obtained with the New Operator and the Kaon-Maid model. In the deuteron, however, the struck nucleon is not at rest but has a finite momentum due to Fermi motion, which smears the relation between the incident photon energy and the invariant mass $W$. 

As illustrated in Fig.~\ref{fig:elementary_dcs}, the elementary cross sections in three of the six relevant isospin channels are significantly larger for the Kaon-Maid model over a broad energy range, while those in the remaining three channels are generally comparable between the two models, except in the case of $\gamma+n\to K^0+\Sigma^0$ channel for $E_\gamma > 1.9$ GeV. The larger elementary cross sections predicted by Kaon-Maid in these three channels therefore suggest that in certain kinematic regions the corresponding nuclear cross section should also be larger than that obtained with the New Operator. The magnitude of the difference at the nuclear level, however, is determined by the combined effects of the elementary amplitudes, Fermi motion, and the nuclear dynamics. Note that, for the $\gamma+n\to K^0+\Lambda$ channel, the cross section shown in Fig.~\ref{fig:elementary_dcs} has been reduced by 50\% for presentation purposes to fit within the plotted scale.

It is also noteworthy that Fig.~\ref{fig:elementary_dcs} indicates measurements of the $\gamma+n\to K^0+\Sigma^0$ channel at energies above 1.9 GeV would be particularly valuable, as the difference between the predictions of the two models begins to increase at this energy. The New Operator was fitted to the limited amount of available data for this channel. Consequently, above 1.9 GeV its cross section represents a prediction rather than an extrapolation constrained by experimental data. Nevertheless, the predicted cross section does not increase dramatically and therefore does not exhibit any divergence.

\begin{widetext}
For the purpose of calculating the observables in kaon photoproduction on the deuteron we may expand the amplitude given in Eq.~(\ref{eq:elementary_ampl_M}) in terms of the complete set of Pauli amplitude \cite{Mart:2008gq,Mart:2025ufa} as
\begin{eqnarray}
\lefteqn{ {t}^{\gamma K}
 ~=~ \left(\frac{E_{N} + m_{N}}{2m_{N}} \right)^{{1}/{2}}
        \left(\frac{E_{Y} + m_{Y}}{2m_{Y}} 
        \right)^{{1}/{2}}}\nonumber\\ && \times\, \chi_{Y}^{\dagger}
        \, \Bigl(\, {\cal{F}}_{1}\,{{\bm \sigma} \,{\bm \cdot}\,} {\bm \epsilon}
+ {\cal{F}}_{2}\,  {\bm{\sigma}}{\bm \,\cdot\,} {\bm k}\, \epsilon_{0}
+ {\cal{F}}_{3}\, {\bm{\sigma}}\bm{\, \cdot \,}{\bm k}\, 
{\bm k}\bm{\, \cdot \,} {\bm{\epsilon}} + {\cal{F}}_{4}\, 
{\bm{\sigma}}\bm{\, \cdot \,}{\bm k}\, {\bm p}_{N}\bm{\, \cdot \,}
{\bm{\epsilon}}  + {\cal{F}}_{5}\, {\bm{\sigma}\,\cdot\,} {\bm k}\, {\bm p}_{Y}\bm{\, \cdot \,}{\bm{\epsilon}}
+ {\cal{F}}_{6}\, {\bm{\sigma}\,\cdot\,} {\bm p}_{N}\,\epsilon_{0} 
+ {\cal{F}}_{7}\, {\bm{\sigma}} 
\cdot {\bm p}_{N}\, {\bm k}\bm{\, \cdot \,}{\bm{\epsilon}}
\nonumber\\ 
&& 
+ {\cal{F}}_{8}\, {\bm{\sigma}}\bm\, \cdot \,{\bm p}_{N}\, {\bm p}_{N} \,{\bm \cdot}\, 
{\bm {\epsilon}} + {\cal{F}}_{9}\, 
{\bm{\sigma}}\bm{\, \cdot \,}{\bm p}_{N}\, 
{\bm p}_{Y}\bm{\, \cdot \,} {\bm{\epsilon}} 
+ {\cal{F}}_{10}\, {\bm{\sigma}}\bm{\, \cdot \,}{\bm p}_{Y}\, 
\epsilon_{0} 
+ {\cal{F}}_{11}\, {\bm{\sigma}}\bm{\, \cdot \,}{\bm p}_{Y}\, {\bm k}\bm{\, \cdot \,}
{\bm{\epsilon}} 
+{\cal{F}}_{12}\, {\bm{\sigma}}
\bm{\, \cdot \,}{\bm p}_{Y}\, {\bm p}_{N}\bm{\, \cdot \,}{\bm{\epsilon}} + 
{\cal{F}}_{13}\, {\bm{\sigma}}\bm{\, \cdot \,}{\bm p}_{Y}\, 
{\bm p}_{Y}\bm{\, \cdot \,} {\bm{\epsilon}}
\nonumber\\ 
&& 
 + {\cal{F}}_{14}\, {\bm{\sigma}}\bm{\, \cdot \,}
{\bm{\epsilon}}\, {\bm{\sigma}}
\bm{\, \cdot \,}{\bm k}\, {\bm{\sigma}}\bm{\, \cdot \,}{\bm p}_{N}+ {\cal{F}}_{15}\, {\bm{\sigma}}\bm{\, \cdot \,}{\bm p}_{Y}\, 
{\bm{\sigma}}\bm{\, \cdot \,}{\bm{\epsilon}}\, {\bm{\sigma}}\bm{\, \cdot \,}{\bm k}
 + {\cal{F}}_{16}\, {\bm{\sigma}}\bm{\, \cdot \,}{\bm p}_{Y}\, 
{\bm{\sigma}}\bm{\, \cdot \,}{\bm{\epsilon}}\, {\bm{\sigma}}\bm{\, \cdot \,}{\bm p}_{N} 
+ {\cal{F}}_{17}\, {\bm{\sigma}}\bm{\, \cdot \,}{\bm p}_{Y}\, 
 {\bm{\sigma}}\bm{\, \cdot \,}
{\bm k}\, {\bm{\sigma}}\bm{\, \cdot \,}{\bm p}_{N}\, \epsilon_{0} 
\nonumber\\ 
 & &  
 + {\cal{F}}_{18}\, {\bm{\sigma}}\bm{\, \cdot \,}{\bm p}_{Y}\, 
{\bm{\sigma}}\bm{\, \cdot \,}{\bm k}\, {\bm{\sigma}\,\cdot\, } 
{\bm p}_{N}\, {\bm k}\bm{\, \cdot \,}{\bm{\epsilon}}
+ {\cal{F}}_{19}\, {\bm{\sigma}}\bm{\, \cdot \,}{\bm p}_{Y}\, 
{\bm{\sigma}}\bm{\, \cdot \,}{\bm k}\, {\bm{\sigma}\,
\cdot\,} {\bm p}_{N}\, {\bm p}_{N}\bm{\, \cdot \,} {\bm{\epsilon}}
 +\, {\cal{F}}_{20}\, {\bm{\sigma}}\bm{\, \cdot \,}{\bm p}_{Y}\, 
{\bm{\sigma}}\bm{\, \cdot \,}{\bm k}\, {\bm{\sigma}} 
\cdot {\bm p}_{N}\, {\bm p}_{Y}\bm{\, \cdot \,} {\bm{\epsilon}}
\, \Bigr) \, \chi_{N} ~.
\label{eq:gen_nonrel_op}
\end{eqnarray}
Note that the amplitudes ${\cal F}_i$ can be expressed in terms of the Lorentz-invariant functions $A_i$ defined in Eq.~(\ref{eq:Ai_Mi}) and listed in Appendix A of Ref.~\cite{Mart:2008gq}. The transition amplitude ${t}^{\gamma K}$ can then be expressed in terms of the spin–non-flip and spin–flip amplitudes, $L$ and $\bm{K}$, respectively, as
\begin{eqnarray}
    \label{eq:L_and_K}
{t}^{\gamma K} = \chi_{Y}^{\dagger}\, \bigl(L + i\,\bm{\sigma}\cdot\bm{K}\bigr)\,\chi_{N},
\end{eqnarray}
with \cite{Mart:1996ay}
\begin{eqnarray}
L & = & N~ \Bigl\{ -\left( {\cal{F}}_{14}+{\cal{F}}_{15}-{\cal{F}}_{16}
\right)~ {\bm p}_N \cdot ({\bm k} \times {\bm{\epsilon}}\, ) 
+{\cal{F}}_{15}~{\bm q}\cdot ({\bm k} \times {\bm{\epsilon}\, })
- {\cal{F}}_{16}~{\bm p}_N \cdot 
({\bm q} \times {\bm{\epsilon}\, })
\nonumber\\ && \hspace{7mm} 
\left.
-\left[ {\cal{F}}_{17}~\epsilon_0 + \left({\cal{F}}_{18}+{\cal{F}}_{20}
\right)~ {\bm k} \cdot {\bm{\epsilon}} \right.
 +\left( {\cal{F}}_{19}+{\cal{F}}_{20} \right)~ {\bm p}_N \cdot 
{\bm{\epsilon}} - {\cal{F}}_{20}~ {\bm q} \cdot {\bm k}
\right]~ {\bm p}_N \cdot ({\bm q} \times {\bm{\epsilon}\, })
\Bigr\} ~,\\
  {\bm K} &=& -N \left( T_1~{\bm{\epsilon}}+T_2~{\bm k}+
T_3~{\bm p}_N+  T_4~{\bm q} \right) ~,
\end{eqnarray}
where
\begin{eqnarray}
\label{eq:normalization}
N &=& \left(\frac{E_{N} + m_{N}}{2m_{N}} \right)^{\frac{1}{2}}
\left(\frac{E_{Y} + m_{Y}}{2m_{Y}} \right)^{\frac{1}{2}} ~ ,
\end{eqnarray}
and
\begin{subequations}
\begin{align}
  T_1 ~&=~ {\cal{F}}_{1}+({\cal{F}}_{14}-{\cal{F}}_{15}-{\cal{F}}_{16})~
          {\bm p}_N \cdot {\bm k}+ {\cal{F}}_{15}~ ({\bm q}\cdot{\bm k}
          - {\bm k}^2) 
          +{\cal{F}}_{16}~({\bm p}_N\cdot{\bm q} - {\bm p}_N^{\, 2}), \\
  T_2 ~&=~ [{\cal{F}}_{2}+{\cal{F}}_{10}+({\bm p}_N\cdot{\bm q}- 
     {\bm p}_N^{\, 2})~{\cal{F}}_{17}]~\epsilon_0+[{\cal{F}}_{3}+{\cal{F}}_{5}
      +{\cal{F}}_{11}+{\cal{F}}_{13}+2{\cal{F}}_{15}+({\bm p}_N \cdot {\bm q}-{\bm p}_N^{\, 2})~({\cal{F}}_{18}+{\cal{F}}_{20})]~{\bm k}\cdot{\bm{\epsilon}} \nonumber\\
      &~~~~ +[{\cal{F}}_{4}+{\cal{F}}_{5}+{\cal{F}}_{12}+{\cal{F}}_{13}-{\cal{F}}_{14}+{\cal{F}}_{15}+{\cal{F}}_{16}+({\bm p}_N \cdot {\bm q}-
      {\bm p}_N^{\, 2})~({\cal{F}}_{19}+
      {\cal{F}}_{20})]~{\bm p}_N\cdot{\bm{\epsilon}}
      -[{\cal{F}}_{5}+{\cal{F}}_{13}+{\cal{F}}_{15}\nonumber\\
      &~~~~ +({\bm p}_N\cdot{\bm q}-{\bm p}_N^{\, 2})~
      {\cal{F}}_{20}]~{\bm q}\cdot{\bm{\epsilon}} ~, \\
  T_3 ~&=~ [{\cal{F}}_{6}+{\cal{F}}_{10}+(2{\bm p}_N\cdot{\bm k}+ 
      {\bm k}^2-{\bm q}\cdot{\bm k})~{\cal{F}}_{17}]~\epsilon_0+
      [{\cal{F}}_{7}+{\cal{F}}_{9}+{\cal{F}}_{11}+{\cal{F}}_{13}
      +{\cal{F}}_{14}+{\cal{F}}_{15}+{\cal{F}}_{16} +(2{\bm p}_N \cdot 
      {\bm k}+{\bm k}^2-{\bm q}\cdot{\bm k})\nonumber\\
      &~~~~ \times ({\cal{F}}_{18}+
      {\cal{F}}_{20})]~{\bm k}\cdot{\bm{\epsilon}}
      + [{\cal{F}}_{8} +{\cal{F}}_{9}+{\cal{F}}_{12}+{\cal{F}}_{13}+2{\cal{F}}_{16}
      +(2{\bm p}_N \cdot {\bm k}+{\bm k}^2-{\bm q}\cdot{\bm k})~
     ({\cal{F}}_{19}+{\cal{F}}_{20})]~{\bm p}_N\cdot
      {\bm{\epsilon}}\nonumber\\
      &~~~~ -[{\cal{F}}_{9}+{\cal{F}}_{13}+{\cal{F}}_{16}
      +(2{\bm p}_N \cdot {\bm k}+{\bm k}^2-{\bm q}\cdot{\bm k})~
      {\cal{F}}_{20}]~{\bm q}\cdot{\bm{\epsilon}} ~,\\
  T_4 ~&=~ -({\cal{F}}_{10}+{\cal{F}}_{17}~{\bm p}_N \cdot {\bm k})~
      \epsilon_0-[{\cal{F}}_{11}+{\cal{F}}_{13}+{\cal{F}}_{15}+
      ({\cal{F}}_{18}+{\cal{F}}_{20})~{\bm p}_N \cdot {\bm k}]~
      {\bm k}\cdot{\bm{\epsilon}}+[{\cal{F}}_{12}+{\cal{F}}_{13}+{\cal{F}}_{16}\nonumber\\
      &~~~~ +
      ({\cal{F}}_{19}+{\cal{F}}_{20})~{\bm p}_N \cdot {\bm k}]~
      {\bm p}_N\cdot{\bm{\epsilon}} +({\cal{F}}_{13}+{\cal{F}}_{20}~
      {\bm p}_N\cdot{\bm k})~{\bm q}\cdot{\bm{\epsilon}} ~.
\end{align}
\end{subequations}
In this form, the spin-independent ($L$) and spin-dependent ($\bm{K}$) parts of the operator are explicitly distinguished, making the representation especially well suited to nuclear applications. Note that this form of the amplitude, given in Eq.~(\ref{eq:L_and_K}), has been used, e.g., in Ref.~\cite{Mart:1996ay} to study hypertriton photoproduction and in Ref.~\cite{Yamamura:1999xm} to examine $\Lambda$- and $\Sigma$-threshold phenomena in inclusive $K^+$ and exclusive $K^+Y$ photoproduction on the deuteron.
\end{widetext}

\subsection{Kaon Photoproduction on the Deuteron}
\begin{figure}[t]
\centering \includegraphics[width=0.4\textwidth]{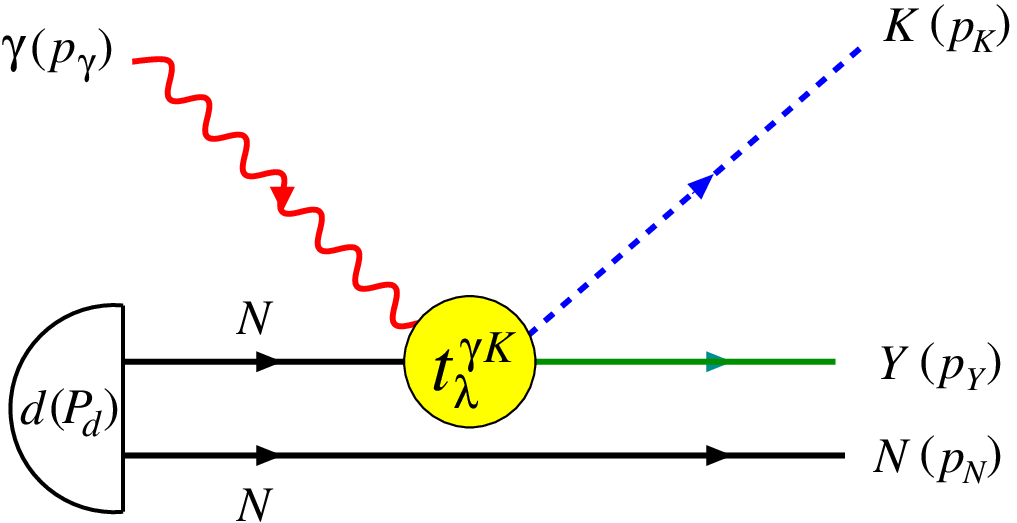} 
\caption{\label{fig:feynman_deut} Feynman diagram for kaon photoproduction on the deuteron, $\gamma d\to KYN$, in the impulse approximation. The corresponding elementary production operator is denoted by $t_\lambda^{\gamma K}$, where $\lambda$ specifies the photon polarization.}
\end{figure}

The formalism for kaon photoproduction on the deuteron was developed in Ref.~\cite{Salam:2004gz} and subsequently refined in Ref.~\cite{Salam:2006kk}. For completeness, we summarize its essential ingredients in a compact form, following the same approach adopted for the elementary production operator. The production process on the deuteron can be written as
\begin{equation}
    \label{eq:deuteron_photo}
    \gamma(p_\gamma)+d(P_d)\to K(p_K)+Y(p_Y)+N(p_N),
\end{equation}
where the four-momenta of the particles, including deuteron, are explicitly written. The corresponding Feynman diagram within the impulse approximation is shown in Fig.~\ref{fig:feynman_deut}. In this approximation, one of the nucleons interacts with the photon to produce a kaon--hyperon pair, while the other nucleon acts as a spectator. It is also important to note that final-state interactions (FSI) are not included in the present treatment. 

\begin{figure}[t]
\centering \includegraphics[width=0.45\textwidth]{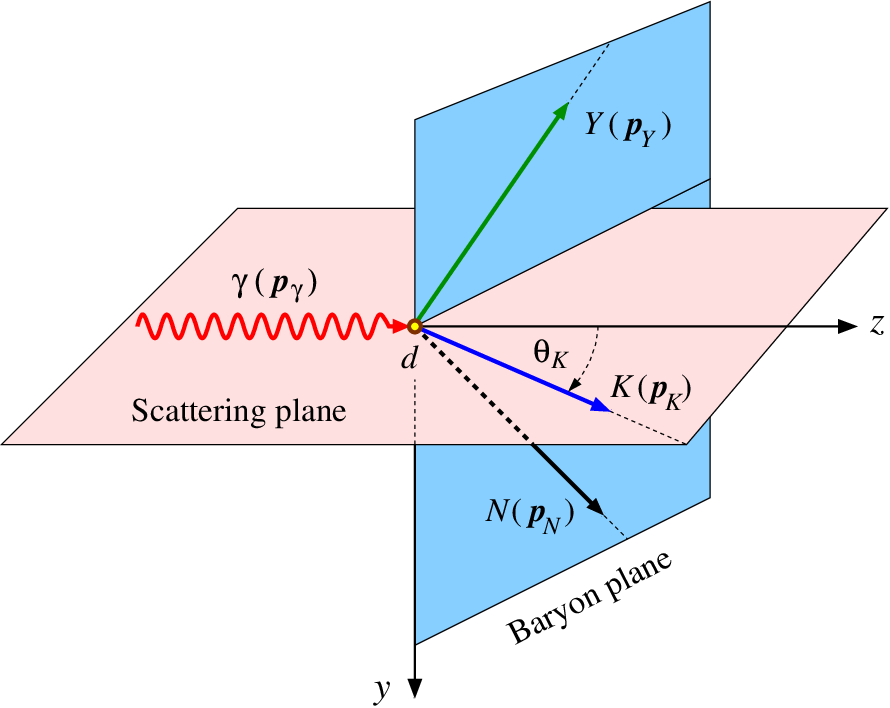} 
\caption{\label{fig:kinematics_dlab} Kinematics of the reaction in the deuteron rest frame. The direction of the incident photon momentum ${\bm p}_{\gamma}$ defines the positive $z$-axis. The kaon momentum ${\bm p}_{K}$ lies in the $xz$-plane, which defines the scattering plane. The positive $y$-axis is defined by ${\bm p}_{\gamma}\times{\bm p}_{K}$. The momenta of the two final-state baryons lie in the baryon plane.}
\end{figure}

The reaction kinematics is illustrated in Fig.~\ref{fig:kinematics_dlab}, with the deuteron taken to be at rest. The photoproduction process can be conveniently described in terms of two planes, i.e., the scattering plane, spanned by the photon and kaon momenta, and the baryon plane, defined by the momenta of the produced nucleon and hyperon. Accordingly, the coordinate system is chosen such that the $y$-axis is perpendicular to the scattering plane, with its direction determined by the cross product of the photon and kaon momenta. With this choice of coordinates and using the four-momentum notation introduced in Eq.~(\ref{eq:deuteron_photo}), the cross section in the deuteron rest frame can be written as
\begin{eqnarray}
d\sigma &=&
\frac{1}{6} \sum_{\alpha\mu_{d}}
\frac{(2\pi)^3}{4E_{\gamma}E_{K}} 
\frac{d{\bm p}_{K}}{(2\pi)^3} \frac{d{\bm p}_{Y}}{(2\pi)^3} 
\frac{d{\bm p}_{N}}{(2\pi)^3} \nonumber \\
&\times & 
\left\vert \sqrt{2}\langle\Psi^{(-)}_{\mu_{Y}\mu_{N}}
\vert t^{\gamma K}_{\lambda}\vert 
\Psi_{\mu_{d}}\rangle\right\vert^2 \nonumber \\
&\times & (2\pi)^{4}\delta^{4}(P_{d}+Q-p_{Y}-p_{N})\,,
\label{eq-gdkyn-dsigma}
\end{eqnarray}
where $\alpha = \{\mu_{Y},\mu_{N},\lambda\}$ denote the spin projections of hyperon, nucleon, and photon 
polarization, respectively, while $\mu_{d}$ is the spin projection of the deuteron. Furthermore, $Q=p_{\gamma}-p_{K}$ is the momentum transfer and the factor $\sqrt{2}$ originates from the proper antisymmetrization of the two-nucleon component of the deuteron wave function.

The inclusive cross section of the process $d(\gamma,K)YN$ is obtained by integrating Eq.~(\ref{eq-gdkyn-dsigma}) in the c.m. frame of the final two baryons as
\begin{eqnarray}
\frac{d\sigma}{dp_{K}d\Omega_{K}} &=&  
\frac{1}{6} \sum_{\alpha}
\int d\Omega^{\rm\,c.m.}_{Y} 
\frac{m_{Y}m_{N}\vert{\bm p}_{K}\vert^2\vert{\bm p}^{\,{\rm c.m.}}_{Y}\vert}
{4(2\pi)^2E_{\gamma}E_{K}W} 
\nonumber\\ &\times&
\left\vert \sqrt{2}\langle\Psi^{(-)}_{\mu_{Y}\mu_{N}}
\vert t^{\gamma K}_{\lambda}\vert
\Psi_{\mu_{d}}\rangle\right\vert^2\,,
\label{eq-gdkyn-dsigma-inclusive}
\end{eqnarray}
where $W^2=(P_{d}+Q)^2$ and $\vert{\bm p}^{\,{\rm c.m.}}_{Y}\vert$ is calculated in 
the center of mass frame of the two final baryons. 

For the polarization observables, we consider only the tensor target asymmetries $T_{2M}$, which are defined by analogy with deuteron photodisintegration~\cite{Are88} such that
\begin{eqnarray}
T_{2M} \frac{d\sigma}{d\Omega_{K}} &=&
(2-\delta_{M0})\, {\mathcal Re}\, V_{2M}\,,\quad M=0, 1, 2\,,
\label{eq-gdkyn-T2m}
\end{eqnarray}
where
\begin{eqnarray}
V_{2M} &=& 
\sqrt{15} 
\sum_{\alpha\mu_{d}\mu^{\prime}_{d}} 
(-1)^{1-\mu^{\prime}_{d}}
\left(\begin{array}{ccc}
 1 & 1 &  2 \\
\mu_{d} & -\mu^{\prime}_{d} & -M 
\end{array}\right) \nonumber\\
&\times&
\int_{0}^{p_K^{\rm max}} d\abs{{\bm p}_{K}} \int d\Omega^{\rm\,c.m.}_{YN}\, \kappa 
{\cal M}_{\alpha\mu_{d}}^{*}
{\cal M}_{\alpha\mu^{\prime}_{d}} ,
\label{eq-gdkyn-V2m}
\end{eqnarray}
with the kinematical factor
\begin{eqnarray}
\kappa &=& \frac{m_{Y} m_{N} \abs{{\bm p}_{K}}^{2}
 \abs{{\bm p}^{\rm\, c.m.}_{YN}}}
{24(2\pi)^{2} E_{\gamma}E_{K} W_{YN}}\,.
\label{eq-gdkyn-kinematical-factor}
\end{eqnarray}
The amplitude ${\cal{M}}_{\alpha\mu_{d}}$ introduced in Eq.~(\ref{eq-gdkyn-V2m}) is given by
\begin{eqnarray}
{\cal{M}}_{\alpha\mu_{d}} &=& \sqrt{2}\langle\Psi_{\mu_{Y}\mu_{N}}^{(-)}
\vert t_{\lambda}^{\gamma K}\vert
\Psi_{\mu_{d}}\rangle ,
\end{eqnarray}
with the $3j$-symbols adhering to the Edmonds convention~\cite{Edm57}. 

\section{Results and Discussion}
\label{sec:Results_and_Discussion}

All results presented in this section are obtained within the impulse approximation, in which the incoming photon interacts with one of the two nucleons in the deuteron while the second nucleon acts as a spectator. The kinematics and the elementary operator are kept in their relativistic form, whereas the deuteron wave function is generated from the Bonn OBEPQ potential and treated non-relativistically. The inclusive cross section of the $d(\gamma,K^0)YN$ reaction is obtained by adding the contributions of the three possible final states, i.e., $K^0\Lambda p$, $K^0\Sigma^0 p$, and $K^0\Sigma^+ n$, the last one originating from the elementary process on the proton. As discussed in Sect.~\ref{sec:intro}, the Neutral Kaon Spectrometer (NKS) measurements were performed at forward kaon angles and close to quasi-free kinematics, a region in which previous studies \cite{Yamamura:1999xm,Salam:2004gz,Miyagawa:2006kj,Salam:2006kk} have consistently shown the final-state interaction and the two-step contributions to be small in comparison with the model dependence of the elementary operator. The impulse approximation therefore provides an adequate framework for the present analysis, in which the elementary operator is the quantity under scrutiny.

\begin{figure*}
\centering\includegraphics[width=0.26\textwidth,angle=-90]{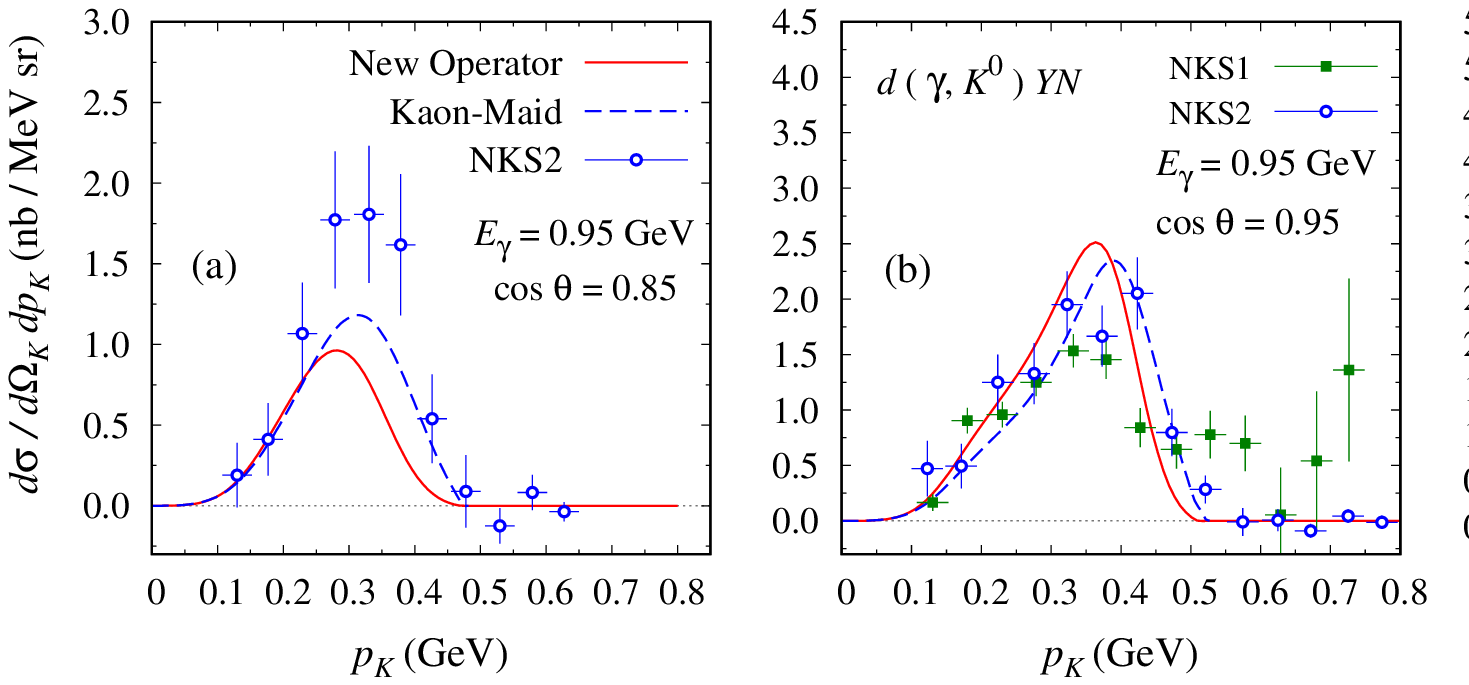} 
\caption{\label{dcs_kmaid_new} Inclusive cross sections for neutral-kaon photoproduction on the deuteron, $d(\gamma,K^0)YN$, as functions of the kaon momentum $p_K$ under different kinematic conditions, specified by $E_\gamma$ and $\cos\theta$. The corresponding kinematics are indicated in each panel. Predictions from Kaon-Maid~\cite{Mart:1999ed,Bennhold:1999mt,kaonmaid} and the New Operator~\cite{Luthfiyah:2021yqe} are compared with experimental data from the NKS1~\cite{Tsukada:2007jy} and NKS2~\cite{Futatsukawa:2012zza} collaborations.}
\end{figure*}

\subsection{Comparison between the Results from Kaon-Maid and the New Operator}

Figure~\ref{dcs_kmaid_new} compares the inclusive cross sections obtained with Kaon-Maid \cite{Mart:1999ed,Bennhold:1999mt,kaonmaid} and with the New Operator \cite{Luthfiyah:2021yqe} at the four kinematic points for which the NKS collaborations have released data, i.e., $E_\gamma = 0.95$ and $1.04$ GeV, each at $\cos\theta = 0.85$ and $0.95$. In all four panels both calculations exhibit a single pronounced peak in the kaon momentum distribution. 

The location of this peak is mainly dictated by the quasi-free kinematics, where the spectator nucleon carries very little momentum and the deuteron wave function is probed at small relative momentum. In this limit, the \(S\)-wave contribution is largest, whereas the \(D\)-wave component decreases as \(p^2\) and vanishes at \(p=0\). Hence, the peak position is primarily a consequence of the reaction kinematics and the low-momentum structure of the deuteron, with only a weak sensitivity to the particular elementary production operator employed.


The magnitude of the peak, in contrast, is controlled by the elementary amplitude evaluated in the region of the invariant mass $W$ selected by the QFS condition and smeared by the Fermi motion of the struck nucleon. As a consequence, the two operators can be cleanly discriminated by the height, rather than by the location, of the quasi-free peak.

The difference between the two predictions grows dramatically with the photon energy. At $E_\gamma = 0.95$ GeV, panels (a) and (b), the two models are still of comparable size, their peaks differ by only about 20\%, with Kaon-Maid slightly above the New Operator at $\cos\theta = 0.85$ and slightly below it at $\cos\theta = 0.95$. Both calculations underestimate the NKS2 points at $\cos\theta = 0.85$, whereas at $\cos\theta = 0.95$ both of them follow the trend of the NKS1 and NKS2 data reasonably well. At $E_\gamma = 1.04$ GeV, panels (c) and (d), the situation changes completely. The Kaon-Maid peak reaches approximately $5.1$ and $7.4$ nb/(MeV sr) at $\cos\theta = 0.85$ and $0.95$, respectively, i.e., more than a factor of two above the corresponding New Operator results of about $2.3$ and $3.5$ nb/(MeV sr), and up to four times larger than the measured cross section. In panel (c) the New Operator reproduces both the position and the height of the observed maximum almost perfectly, while Kaon-Maid overshoots the data by more than a factor of two.

This behavior can be traced back directly to the elementary process displayed in Fig.~\ref{fig:elementary_dcs}. The inclusive $d(\gamma,K^0)YN$ cross section in this energy region is dominated by the $\gamma + n \to K^0 + \Lambda$ channel, for which Kaon-Maid was never constrained by data. Its prediction for this channel is known to be unrealistically large \cite{Salam:2004gz}; indeed, in Fig.~\ref{fig:elementary_dcs} the Kaon-Maid curve for $\gamma + n \to K^0 + \Lambda$ had to be scaled by a factor of $0.5$ in order to fit within the plotted range. Since the New Operator was fitted to the recent CLAS and MAMI data in this channel \cite{CLAS:2017gsu,A2:2018doh}, its nuclear prediction inherits a far more reliable elementary input. Furthermore, increasing $E_\gamma$ from $0.95$ to $1.04$ GeV shifts the relevant elementary invariant mass into the region $W \approx 1.7$--$1.8$ GeV, precisely where the two elementary models diverge most strongly. This explains why the discrepancy between the two nuclear calculations is modest at the lower photon energy and becomes severe at the higher one. We therefore conclude that the New Operator provides a substantially better description of the inclusive deuteron data than Kaon-Maid, and that the deuteron cross section at $E_\gamma \gtrsim 1$ GeV is a sensitive discriminator between elementary models.

\begin{figure*}
\centering\includegraphics[width=0.4\textwidth,angle=-90]{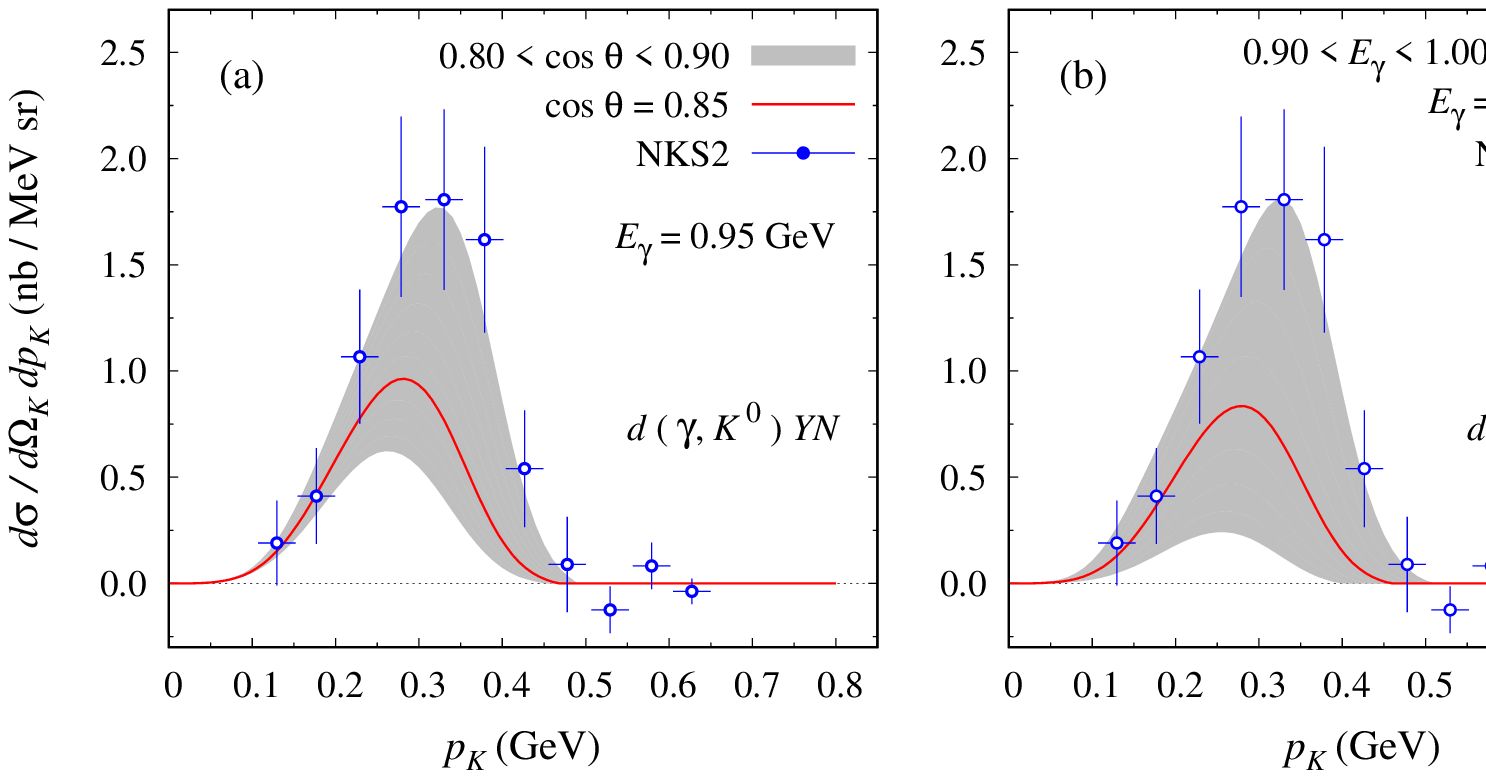} 
\caption{\label{dcs_e095_cth085} Inclusive cross section for neutral-kaon photoproduction on the deuteron, $d(\gamma,K^0)YN$, as a function of the kaon momentum $p_K$ for $0.90<E_\gamma<1.00$ GeV and $0.80<\cos\theta<0.90$, compared with experimental data from NKS2 \cite{Futatsukawa:2012zza}. The uncertainty in the calculated cross section due to the finite kaon angular bin is shown by the shaded gray area in panel (a), while the corresponding uncertainty due to the photon-energy bin is shown in panel (b). The cross sections evaluated at $\cos\theta=0.85$ and $E_\gamma=0.95$ GeV are shown by the solid red lines. Note that all calculations are performed by using the new elementary operator \cite{Luthfiyah:2021yqe}.}
\end{figure*}

\begin{figure*}
\centering \includegraphics[width=0.4\textwidth,angle=-90]{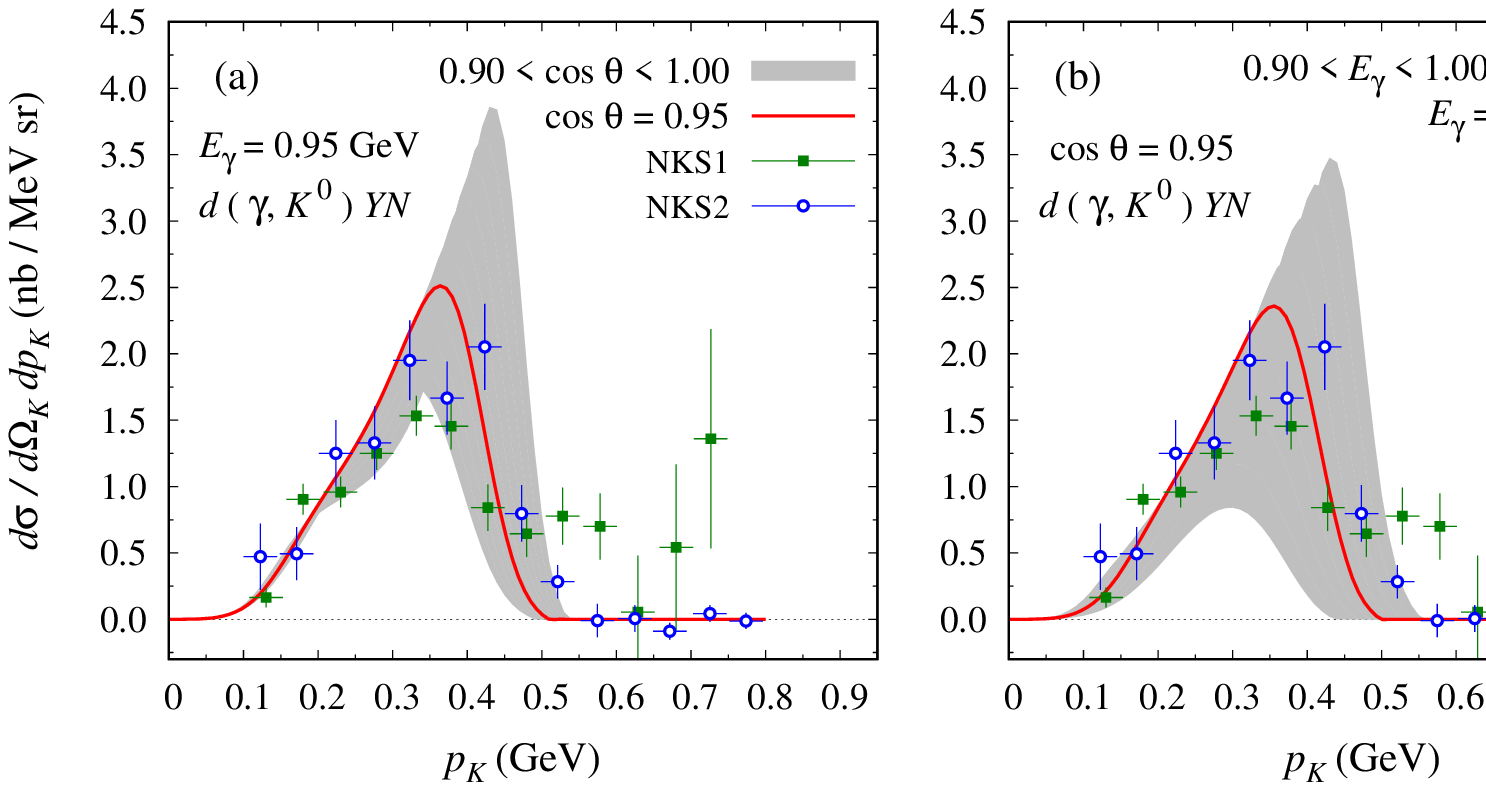} 
\caption{\label{dcs_e095_cth095} As in Fig.~\ref{dcs_e095_cth085}, but for $0.90<E_\gamma<1.00$ GeV and $0.90<\cos\theta<1.00$. Experimental data are taken from NKS1 \cite{Tsukada:2007jy} and NKS2 \cite{Futatsukawa:2012zza}.}  
\end{figure*}

\begin{figure*}
\centering \includegraphics[width=0.4\textwidth,angle=-90]{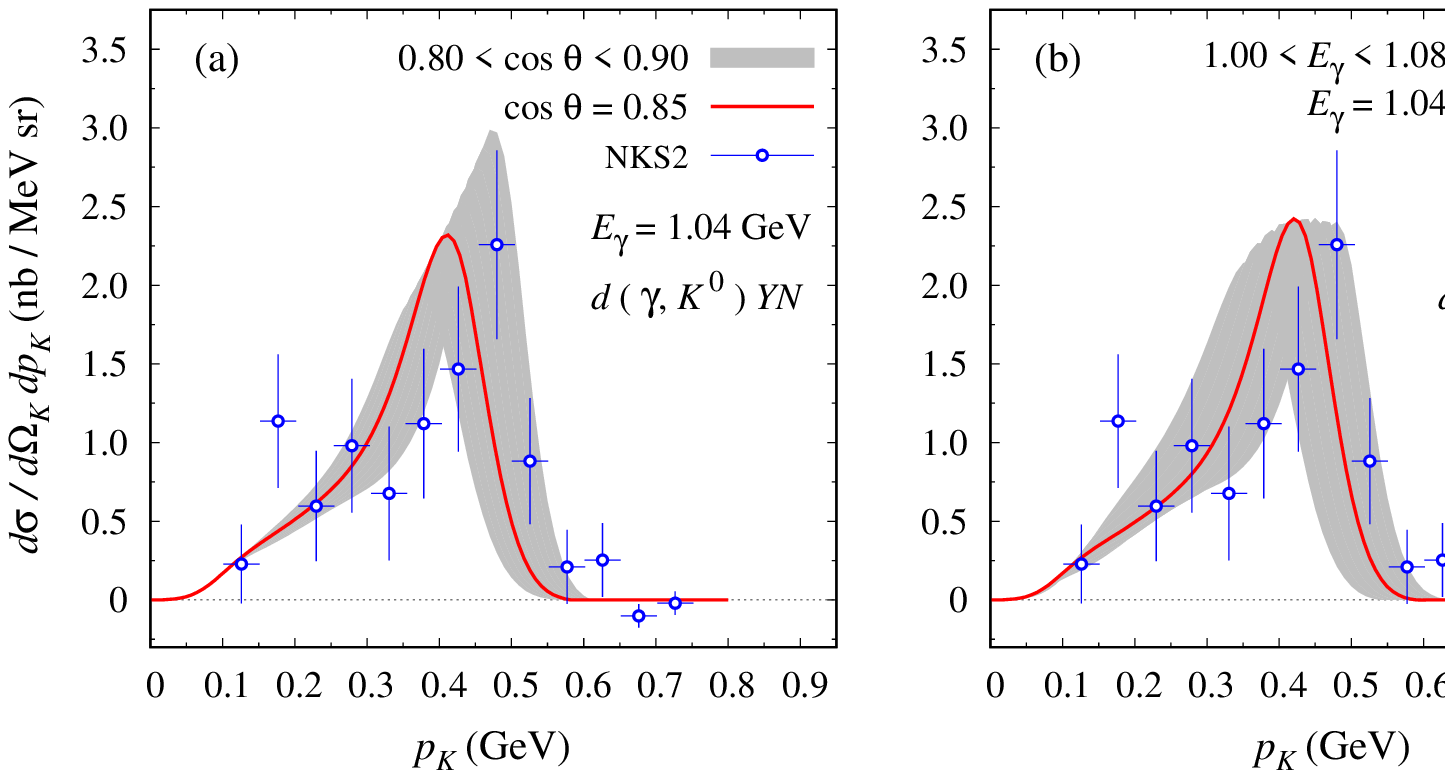} 
\caption{\label{dcs_e100_cth085} As in Fig.~\ref{dcs_e095_cth085}, but for $1.00<E_\gamma<1.08$ GeV and $0.80<\cos\theta<0.90$. Experimental data are taken from NKS2 \cite{Futatsukawa:2012zza}.}  
\end{figure*}

\begin{figure*}
\centering \includegraphics[width=0.4\textwidth,angle=-90]{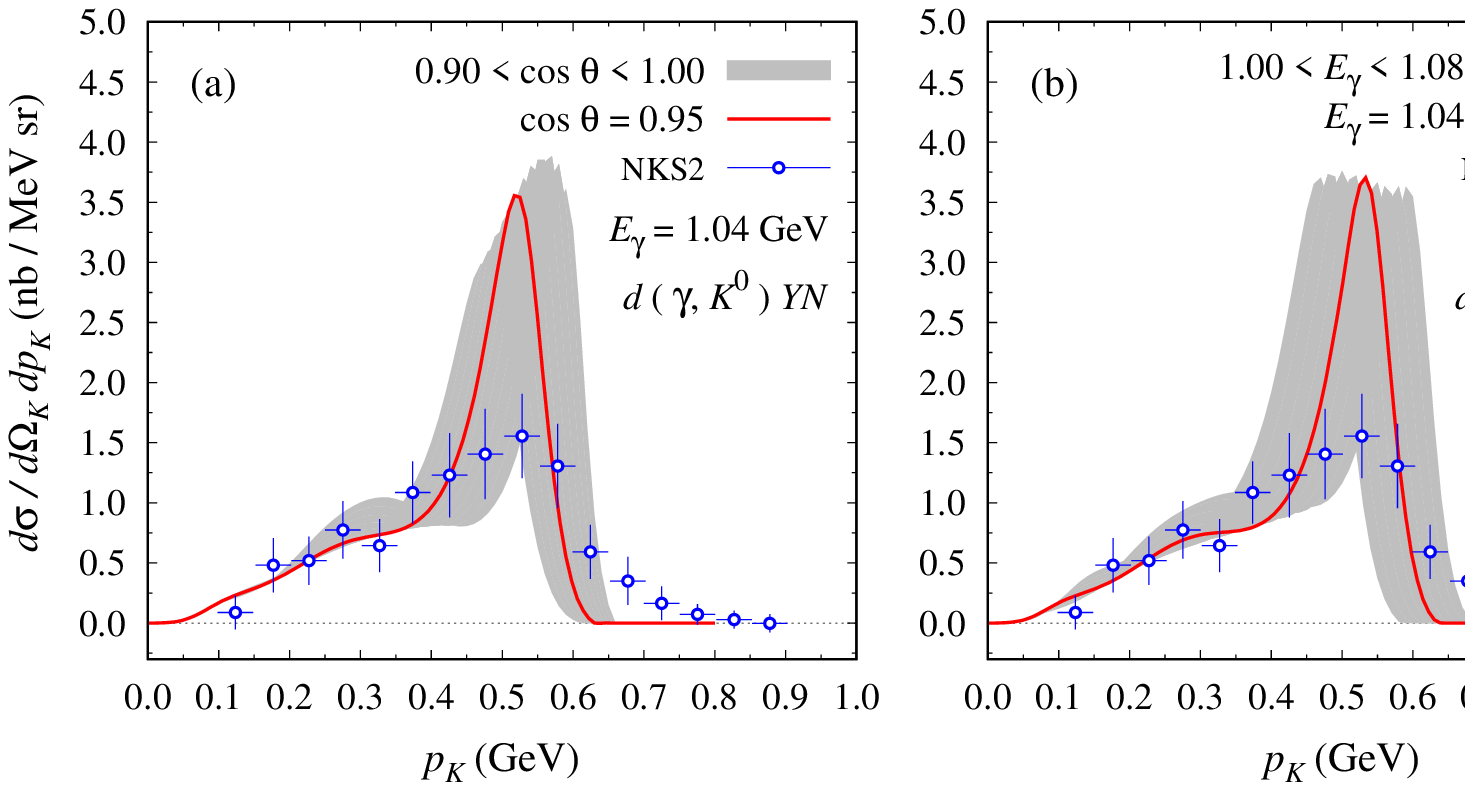} 
\caption{\label{dcs_e100_cth095}As in Fig.~\ref{dcs_e095_cth085}, but for $1.00<E_\gamma<1.08$ GeV and $0.90<\cos\theta<1.00$. Experimental data are taken from NKS2 \cite{Futatsukawa:2012zza}.}  
\end{figure*}

\subsection{Effects of the finite energy and angular bins}

The comparison presented in Fig.~\ref{dcs_kmaid_new} is performed at a single photon energy and a single kaon angle, whereas the experimental points are averaged over bins of finite width, typically $\Delta\cos\theta = 0.10$ and $\Delta E_\gamma \approx 80$--$100$ MeV. Because the position of the quasi-free peak depends strongly on both variables, this averaging is far from innocuous. To quantify it we have recalculated the cross section, using the New Operator throughout, over the full width of the experimental bins. The results are collected in Figs.~\ref{dcs_e095_cth085}--\ref{dcs_e100_cth095}, where the shaded gray areas display the range spanned by the calculation within the angular bin, panel (a), and within the photon-energy bin, panel (b), while the solid lines show the cross sections evaluated at the central values.

Two observations follow immediately. First, the two sources of uncertainty are comparable in size, since both act through the same QFS relation between $p_K$, $E_\gamma$, and $\cos\theta$. Second, the resulting band is much wider than one might naively expect, i.e., near the maximum, the upper edge of the band can exceed the central value by roughly a factor of two. The consequence for the interpretation of the data is important. The apparent underestimation of the NKS2 points seen in Fig.~\ref{dcs_kmaid_new}(a) is entirely removed once the bin width is taken into account, as demonstrated in Fig.~\ref{dcs_e095_cth085}, where the measured points lie comfortably inside the shaded band. The same holds in Fig.~\ref{dcs_e095_cth095} for the more forward bin $0.90 < \cos\theta < 1.00$, and in Fig.~\ref{dcs_e100_cth085} at the higher photon energy, where the central curve already describes the data very well.

The most instructive case is Fig.~\ref{dcs_e100_cth095}, corresponding to $1.00 < E_\gamma < 1.08$ GeV and $0.90 < \cos\theta < 1.00$. Here the central calculation overshoots the measured maximum by about a factor of two, but the data still fall within the shaded band, whose lower edge is strongly suppressed because a substantial part of the bin corresponds to kinematics far from the QFS condition. The band is also markedly asymmetric, with its width increasing toward the forward direction and higher photon energies. This behavior follows from the kinematic mapping $p_K=p_K(E_\gamma,\cos\theta)$, since in these regions, the derivatives $\left|\partial p_K/\partial\cos\theta\right|$ and $\left|\partial p_K/\partial E_\gamma\right|$ become larger, so finite experimental energy and angular bins translate into a larger spread in $p_K$. This observation carries a clear practical message, i.e., a meaningful comparison between theory and experiment in this reaction requires the calculation to be properly averaged over the finite experimental energy and angular bins, and any conclusion drawn from a single-point evaluation should therefore be treated with caution. Conversely, future measurements performed with narrower energy and angular bins would reduce this ambiguity considerably and sharpen the discrimination between the elementary operators discussed in the previous subsection.

We note, finally, that the NKS1 points at large kaon momenta, $p_K \gtrsim 0.55$ GeV in Fig.~\ref{dcs_e095_cth095}, are not reproduced by the calculation, which drops to zero once the kinematic boundary of the quasi-free mechanism is passed. Such strength cannot be generated by the impulse approximation and lies outside the band obtained from the bin averaging. Whether this excess strength results from experimental momentum resolution and the associated bin migration, or from reaction mechanisms not included in the present framework, cannot be determined from the currently available data. It is also worth noting that, in this kinematic region, the NKS1 and NKS2 data exhibit a significant discrepancy.

\begin{figure}
\centering \includegraphics[width=0.45\textwidth]{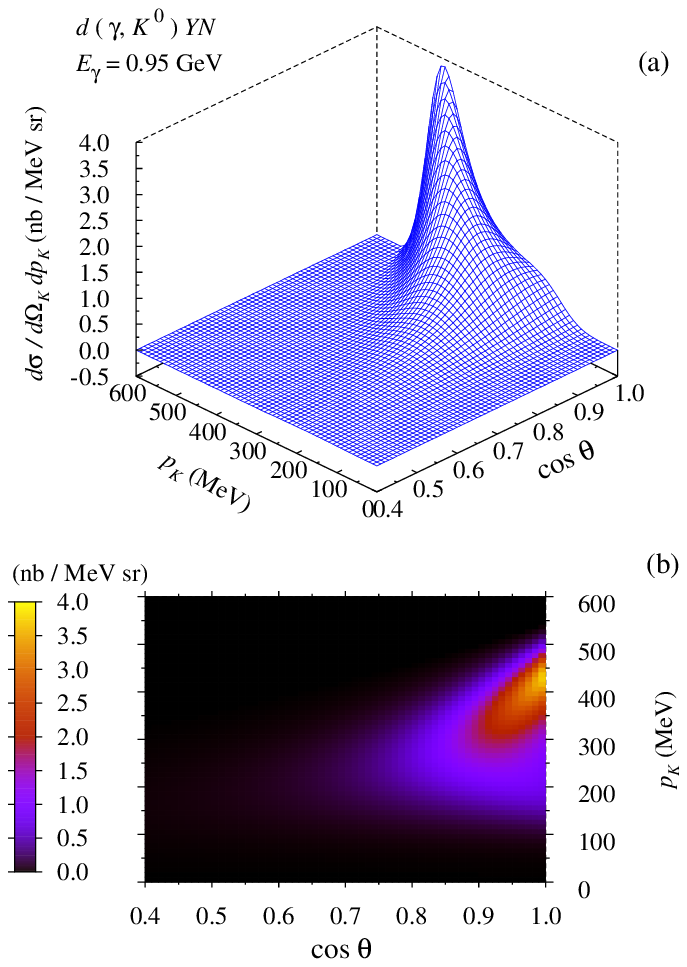} 
\caption{\label{dcs_3D_950} (a) Three-dimensional plot of the inclusive differential cross section as a function of the kaon momentum $p_K$ and the cosine of the kaon emission angle, $\cos\theta$, at $E_\gamma = 0.95$ GeV. (b) Corresponding projection of the cross section onto the $p_K$--$\cos\theta$ plane, showing the location of the maximum and the direction of the ridge extending from the peak toward smaller $p_K$ as the kaon emission angle moves away from the forward direction.}
\end{figure}

\begin{figure}
\centering \includegraphics[width=0.45\textwidth]{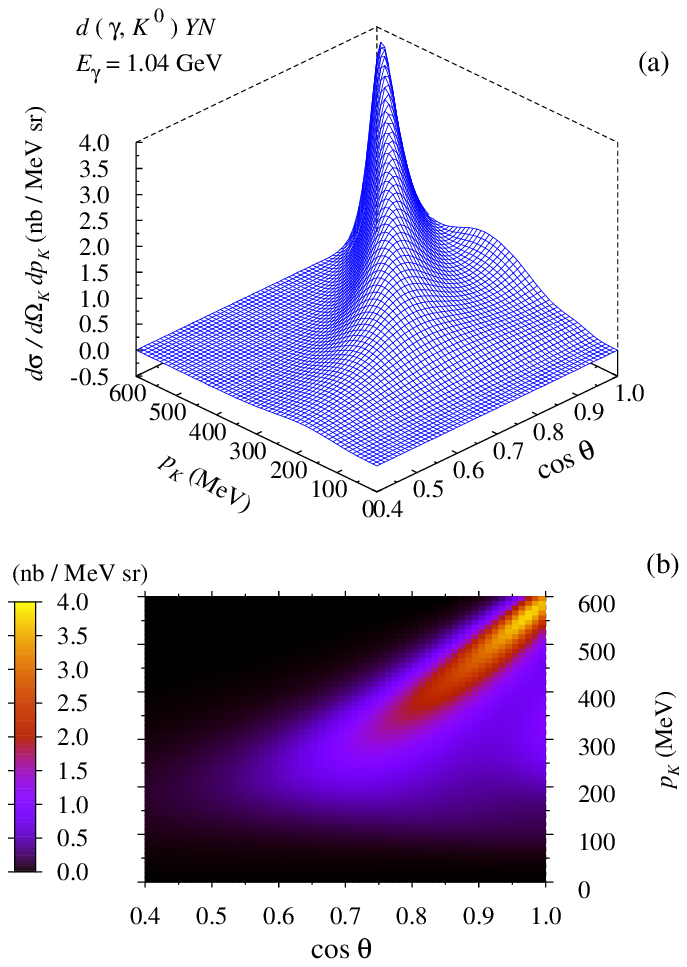} 
\caption{\label{dcs_3D_1040} As in Fig.~\ref{dcs_3D_950}, but for $E_\gamma = 1.04$ GeV.}  
\end{figure}

\begin{figure}
\centering \includegraphics[width=0.45\textwidth]{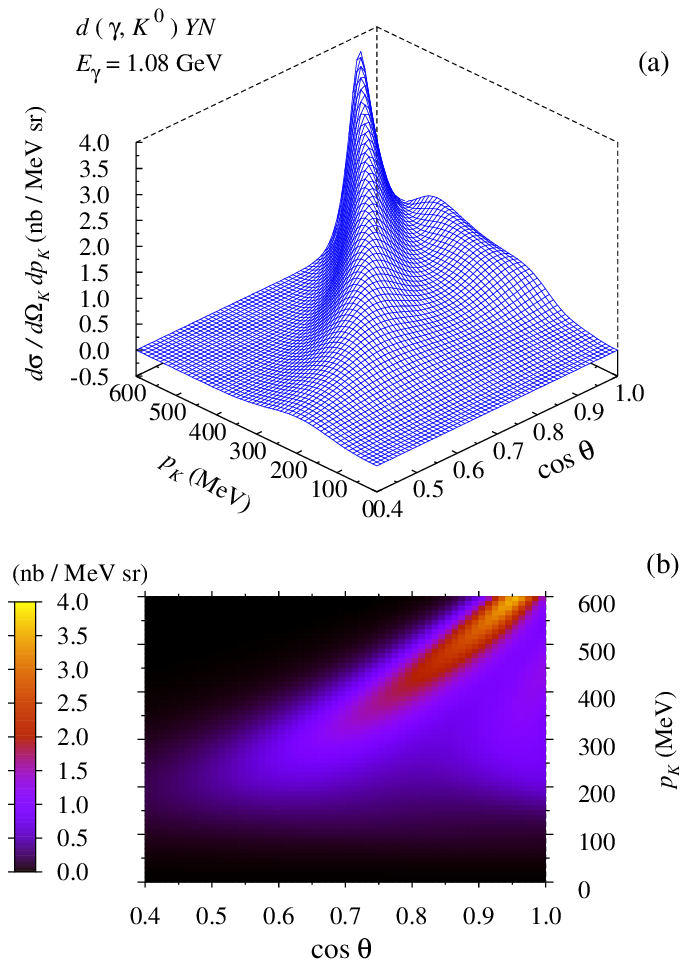} 
\caption{\label{dcs_3D_1080} As in Fig.~\ref{dcs_3D_950}, but for $E_\gamma = 1.08$ GeV.} 
\end{figure}

\subsection{Global view of the cross section and optimal kinematics}

In order to obtain a more comprehensive picture and, at the same time, to provide guidance for future experiments, we display in Figs.~\ref{dcs_3D_950}--\ref{dcs_3D_1080} the inclusive cross section as a three-dimensional surface over the $p_K$--$\cos\theta$ plane, together with its projection onto that plane, for $E_\gamma = 0.95$, $1.04$, and $1.08$ GeV. All three figures are obtained by using the New Operator.

The dominant feature in all cases is a narrow ridge that runs along the quasi-free locus. The ridge attains its maximum in the most forward direction, $\cos\theta \to 1$, and moves steadily toward smaller kaon momenta as the kaon emission angle increases, while its height decreases rapidly. The position of the maximum shifts with the photon energy, from $p_K \approx 0.4$ GeV at $E_\gamma = 0.95$ GeV to $p_K \approx 0.5$ and $0.55$ GeV at $E_\gamma = 1.04$ and $1.08$ GeV, respectively, in accordance with the QFS condition on a nucleon at rest. Away from the ridge the cross section falls by more than an order of magnitude, because the spectator nucleon is then forced to carry a large momentum, for which the deuteron wave function is strongly suppressed.

A second, considerably broader structure appears at smaller kaon momenta. It is barely visible at $E_\gamma = 0.95$ GeV, develops into a shoulder at $E_\gamma = 1.04$ GeV, and becomes a clearly separated bump at $E_\gamma = 1.08$ GeV, where it extends over the region $\cos\theta \approx 0.6$--$0.9$. This evolution is naturally explained by the opening of the $\Sigma$ channels: the free thresholds of the $\gamma n \to K^0\Sigma^0$ and $\gamma p \to K^0\Sigma^+$ reactions lie slightly above $E_\gamma \approx 1.05$ GeV, so that at $E_\gamma = 1.04$ GeV these channels can be reached only sub-threshold, through the Fermi motion of the struck nucleon, whereas at $E_\gamma = 1.08$ GeV they begin to open. The inclusive cross section in this energy range thus separates, in the $p_K$--$\cos\theta$ plane, into a $\Lambda$ quasi-free ridge at large $p_K$ and a $\Sigma$ quasi-free structure at smaller $p_K$.

These maps suggest a concrete strategy for future measurements. Experiments aiming at the extraction of the elementary $\gamma n \to K^0\Lambda$ amplitude, or at the discrimination between elementary operators, should concentrate on $\cos\theta > 0.9$ and on a narrow window of kaon momenta centered on the ridge, since this is simultaneously the region of the largest counting rate and the region in which the model dependence is most pronounced, as seen in Fig.~\ref{dcs_kmaid_new}. Because the ridge is narrow, however, such a measurement is only meaningful if the angular and energy bins are kept small; the bands of Figs.~\ref{dcs_e095_cth085}--\ref{dcs_e100_cth095} show what is lost otherwise. Conversely, an experiment aiming at the $\Sigma$ channels should be performed at $E_\gamma \gtrsim 1.08$ GeV and at moderately forward angles, $\cos\theta \approx 0.7$--$0.9$, where the $\Sigma$ structure is best separated from the dominant $\Lambda$ ridge.

\begin{figure}
\centering \includegraphics[width=0.45\textwidth]{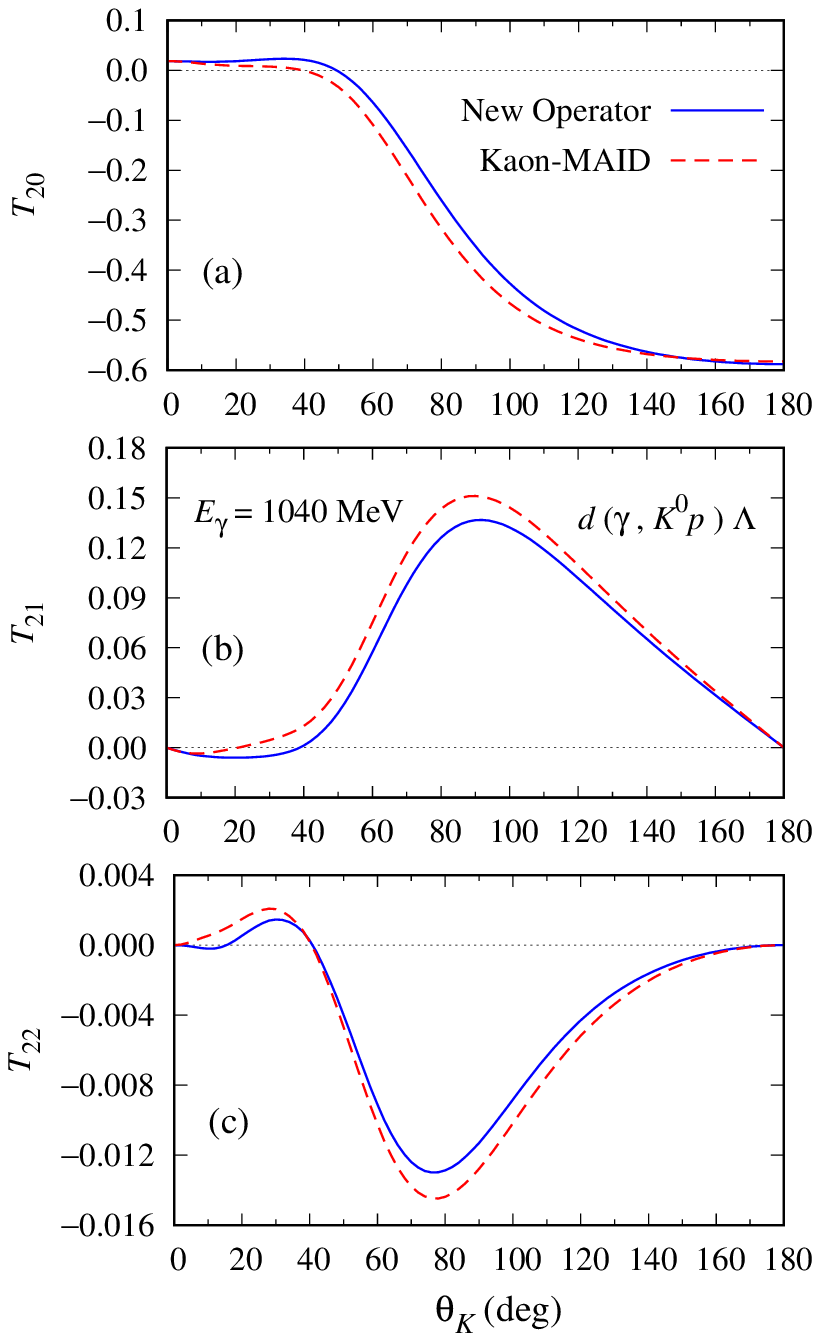} 
\caption{\label{tensor_1040} Tensor target asymmetries ${T}_{20}$, ${T}_{21}$, and ${T}_{22}$, for the inclusive neutral-kaon photoproduction on the deuteron, $d(\gamma,K^0)YN$, as functions of the kaon angle $\theta_K$ at $E_\gamma=1040$ MeV. The solid and dashed lines correspond to the results obtained with the New Operator~\cite{Luthfiyah:2021yqe} and with Kaon-Maid~\cite{Mart:1999ed,Bennhold:1999mt,kaonmaid}, respectively. The asymmetry is defined in Ref.~\cite{Salam:2006kk}.}
\end{figure}

\subsection{Tensor target asymmetries}

Finally, we turn to the polarization observables. Figure~\ref{tensor_1040} shows the tensor target asymmetries ${T}_{20}$, ${T}_{21}$, and ${T}_{22}$, defined as in Ref.~\cite{Salam:2004gz}, as functions of the kaon angle $\theta_K$ at $E_\gamma = 1040$ MeV, calculated with both elementary operators. The three asymmetries display distinctly different angular patterns. The asymmetry ${T}_{20}$ is essentially zero for $\theta_K \lesssim 40^\circ$ and then decreases monotonically, reaching about $-0.6$ in the backward direction. The asymmetry ${T}_{21}$ exhibits a broad positive maximum of about $0.14$ near $\theta_K \approx 90^\circ$ and vanishes at both ends of the angular range, while ${T}_{22}$ remains very small over the whole range, with a shallow negative minimum of about $-0.013$ around $\theta_K \approx 80^\circ$.

The most striking feature of Fig.~\ref{tensor_1040} is that the two elementary operators yield remarkably similar predictions, in sharp contrast to the situation found for the cross section. Kaon-Maid produces a slightly faster decrease of ${T}_{20}$ between $40^\circ$ and $140^\circ$, a maximum of ${T}_{21}$ that is larger by about $10\%$, and a slightly deeper minimum of ${T}_{22}$, but the shapes of all three observables are practically unchanged. This is easily understood, since the tensor asymmetries are ratios of bilinear combinations of the same amplitudes, so that the overall normalization of the elementary operator, which is what distinguishes the two models most strongly, largely cancels. The asymmetries are therefore governed primarily by the spin structure of the deuteron and by the reaction mechanism rather than by the details of the elementary amplitude. They constitute a valuable complementary observable, in the sense that they test the nuclear part of the calculation in a way that the cross section cannot.

From the experimental point of view, however, these observables are demanding. The region in which ${T}_{20}$ becomes sizable is the backward hemisphere, where the inclusive cross section is smallest, and ${T}_{22}$ is of the order of $10^{-2}$ throughout. The most promising compromise is offered by ${T}_{21}$ in the intermediate range $\theta_K \approx 60^\circ$--$100^\circ$, where the asymmetry is of the order of $0.1$ and the cross section, although reduced, is not yet negligible. It should also be kept in mind that at the larger angles the final-state interaction and the pion-mediated two-step process are known to become significant \cite{Salam:2004gz,Miyagawa:2006kj}, so that a quantitative analysis of the backward region would require these mechanisms to be included explicitly.

\section{Conclusion}
\label{sec:Conclusion}

We have investigated the inclusive photoproduction of neutral kaons on the deuteron, $d(\gamma,K^0)YN$, for photon energies between $0.9$ and $1.1$ GeV, using the impulse approximation together with a deuteron wave function generated from the Bonn OBEPQ potential and, as elementary input, the New Operator introduced in Ref.~\cite{Luthfiyah:2021yqe}.

The comparison with the Kaon-Maid model shows that the choice of the elementary operator has a significant influence on the nuclear observables. Whereas the position of the quasi-free peak is fixed by kinematics and is therefore common to both models, its magnitude differs by more than a factor of two at $E_\gamma = 1.04$ GeV. The New Operator describes the NKS1 and NKS2 data considerably better than Kaon-Maid, which overestimates the measured cross section by up to a factor of four. This difference originates from the elementary $\gamma n \to K^0\Lambda$ channel, for which Kaon-Maid was never constrained by experimental data and consequently yields an unrealistically large cross section, whereas the New Operator has been fitted to a modern and comprehensive database.

We have further shown that the finite photon-energy and kaon-angle bins of the existing measurements generate a theoretical uncertainty that is comparable to, and in the forward direction even larger than, the difference between the models themselves. Once the calculation is averaged over the experimental bins, the New Operator results are consistent with all available NKS1 and NKS2 points. This implies that a single-point evaluation of the cross section is not sufficient for a quantitative confrontation with these data, and that a genuine improvement of our knowledge of the elementary neutral-kaon amplitudes requires measurements with substantially narrower bins.

The three-dimensional maps of the cross section reveal a narrow quasi-free ridge that dominates the $p_K$--$\cos\theta$ plane and moves toward smaller kaon momenta as the emission angle grows, accompanied by a second, broader structure that emerges at $E_\gamma \gtrsim 1.04$ GeV and is associated with the opening of the $\Sigma$ channels. These maps identify the most favorable kinematics for future experiments, i.e., $\cos\theta > 0.9$ and a narrow momentum window around the ridge for studies of the $\Lambda$ channel and of the elementary operator, and $E_\gamma \gtrsim 1.08$ GeV with $\cos\theta \approx 0.7$--$0.9$ for the $\Sigma$ channels. Within these quasi-free kinematic regions, future theoretical investigations of polarization and other reaction observables, as well as studies employing different theoretical ingredients, can be pursued in kinematic settings that remain sufficiently accessible for experimental measurements.

These results also point to the possibility of extracting the elementary  $\gamma N\to K^0Y$ amplitudes from deuteron data once sufficiently precise measurements become available in the identified quasi-free regions. There, the dominance of the quasi-free mechanism provides a particularly clean connection between the measured nuclear observables and the underlying elementary amplitudes, while measurements with narrow energy, angular, and momentum bins would substantially reduce the kinematic smearing discussed above. Such an extraction would provide an important test of the elementary operator in a channel for which existing data are scarce and model dependence remains substantial. Developing a quantitative procedure for this extraction, including a controlled treatment of the remaining nuclear effects, is one of the directions we plan to pursue in our future work.

Finally, the tensor target asymmetries ${T}_{20}$, ${T}_{21}$, and ${T}_{22}$ have been found to be far less sensitive to the elementary operator than the cross section, since the normalization of the elementary amplitude largely cancels in these ratios. They therefore probe the nuclear dynamics rather than the elementary input and provide information complementary to that contained in the cross section. Their measurement is nevertheless challenging, since the asymmetries are largest in the angular region where the cross section is smallest and where final-state interactions can no longer be neglected. As a consequence, an extension of the present framework to include $YN$ and $KN$ rescattering, as well as the pion-mediated process, following the approach of Refs.~\cite{Salam:2004gz,Salam:2006kk}, provides an opportunity to further investigate these observables and the underlying nuclear dynamics.

\section*{Acknowledgements}
This work was supported by the PUTI Q1 Grant from University of Indonesia under contract PKS-206/UN2.RST/HKP.05.00/2025. The authors acknowledge that AI tools (Claude and ChatGPT) were used in portions of this article to assist with English-language polishing and to improve the clarity and presentation of the text. Claude was also used to help sharpen the authors’ analysis of the numerical results. However, the scientific content, final analysis and interpretation presented in this article are entirely the authors’ own.

\bibliographystyle{elsarticle-num}
\bibliography{ref}

\end{document}